\documentclass[twocolumn]{article}
\usepackage{graphicx}
\usepackage{xspace}
\usepackage{amsmath}
\usepackage{amssymb}
\usepackage[usenames]{color}

\newcommand{\dd}{\mathrm{d}}

\newcommand{\eg}{e.g.~}
\newcommand{\eq}[1]{Eq.~\ref{eq:#1}}
\newcommand{\eqs}[1]{Eqs.~\ref{eq:#1}}
\newcommand{\fig}[1]{Fig.~\ref{fig:#1}}
\newcommand{\figs}[1]{Figs.~\ref{fig:#1}}
\newcommand{\sect}[1]{Section \ref{sec:#1}}

\newcommand{\tabl}[1]{Table \ref{table:#1}}

\newcommand{\nba}[1]{} 
\newcommand{\mAA}{{\text \AA}\xspace}

\newcommand{\qmax}{\ensuremath{Q_{\mathrm{max}}}\xspace}
\newcommand{\qmin}{\ensuremath{Q_{\mathrm{min}}}\xspace}
\newcommand{\rmin}{\ensuremath{r_{\mathrm{min}}}\xspace}
\newcommand{\rmax}{\ensuremath{r_{\mathrm{max}}}\xspace}

\newcommand{\mmax}{\ensuremath{m_{\mathrm{max}}}\xspace}
\newcommand{\delr}{\ensuremath{dr}\xspace}
\newcommand{\delrn}{\ensuremath{dr_N}\xspace}

\newcommand{\qualp}{\ensuremath{\mathcal{Q}_p}\xspace}

\newcommand{\PDFgui}{\textsc{PDFgui}\xspace}
\newcommand{\FittwoD}{\textsc{Fit2D}\xspace}
\newcommand{\PDFgetXtwo}{\textsc{PDFgetX2}\xspace}
\newcommand{\PDFgetN}{\textsc{PDFgetN}\xspace}

\begin{document}

\title{The Nyquist-Shannon sampling theorem and the atomic pair distribution function}

\author{
Christopher L. Farrow\\
\small Department of Applied Physics and Applied Mathematics, Columbia University,\\ 
\small New York, NY, 10027, USA \\
\and
Margaret Shaw\\
\small Department of Physics and Astronomy, Princeton University \\
\small Princeton, New Jersey, 08544, USA \\
\and
Hyunjeong Kim\\
\small Energy Technology Research Institute, National Institute of Advanced Industrial Science and Technology \\
\small Tsukuba, Ibaraki, 305-8565, Japan \\
\and
Pavol Juh\'{a}s\\
\small Department of Applied Physics and Applied Mathematics, Columbia University,\\ 
\small New York, NY, 10027, USA\\
\and
Simon J. L. Billinge\\
\small Department of Applied Physics and Applied Mathematics, Columbia University,\\ 
\small New York, NY, 10027, USA \\
\small Condensed Matter Physics and Materials Science Department, Brookhaven
National Laboratory,\\
\small Upton, New York, 11973, USA\\
\small \texttt sb2896@columbia.edu
}

\date{}

\maketitle

\begin{abstract}

We have systematically studied the optimal real-space sampling of atomic pair
distribution data by comparing refinement results from oversampled and
resampled data. Based on nickel and a complex perovskite system, we demonstrate
that the optimal sampling is bounded by the Nyquist interval described by the
Nyquist-Shannon sampling theorem. Near this sampling interval, the data points
in the PDF are minimally correlated, which results in more reliable uncertainty
prediction.  Furthermore, refinements using sparsely sampled data may run many
times faster than using oversampled data. This investigation establishes a
theoretically sound limit on the amount of information contained in the PDF,
which has ramifications towards how PDF data are modeled.

\end{abstract}

\section{Introduction}

Atomic pair distribution function (PDF) analysis of x-ray and neutron powder
diffraction data is becoming prominent in structure analysis of complex
materials due to an increasing interest in studying structure from nanoscale
structural order.~\cite{billi;s07}  Dedicated experimental facilities are
appearing for PDF studies~\cite{proff;apa01i,chupa;jac07} as well as
specialized
software.~\cite{farro;jpcm07,qiu;jac04i,tucke;jac01a,peter;jac00,proff;jac99} As
more people in the structure-characterization community adopt the PDF method,
it is important to reevaluate and strengthen our analysis techniques.  To this
end, we have investigated the information content in the PDF data allowing us
to determine optimal grid spacings to use when calculating PDFs.  The sampling
grid for PDFs is typically chosen in an \emph{ad-hoc} way, for example, to give
a visually smooth PDF. The information content in the PDF does not increase for
grid intervals above a critical value.  If the data are oversampled, not only
is no new information introduced, the points in the PDF are not statistically
independent,~\cite{toby;aca04,egami;b;utbp03} which leads to improper estimates
of uncertainties in refinement parameters and slowing down the
refinement.~\cite{schwa;aca89}

We have systematically studied the optimal PDF sampling interval for PDF data
and demonstrate that it is consistent with the value predicted by the
Nyquist-Shannon sampling theorem.~\cite{shann;pire49} This gives the minimum
amount of information we need to completely specify a PDF from a given $F(Q)$.
When this optimal sampling is enforced, we see significant speed-up in our PDF
refinements accompanied by a small increase in estimated uncertainties due to
the reduction of statistical correlations among the PDF points.  When the data
are made sparser than the optimal sampling interval the refinement results
rapidly become unreliable due to aliasing.

\section{The PDF method}

The PDF method is a total scattering technique for determining local order in
nanostructured materials.~\cite{egami;b;utbp03} The technique does not require
periodicity, so it is well suited for studying nanoscale features in a variety
of materials.~\cite{billi;jssc08,billi;cc04} The experimental PDF, denoted
$G(r)$, is the truncated Fourier transform of the total scattering structure
function, $F(Q) = Q[S(Q) - 1]$:~\cite{farro;aca09}
\begin{equation}
\label{eq:FTofSQtoGr}
  G(r) = \frac{2}{\pi}
          \int_{\qmin}^{\qmax} F(Q)\sin(Qr) \: \dd Q,
\end{equation}
where $Q$ is the magnitude of the scattering momentum. The structure function,
$S(Q)$, is extracted from the Bragg and diffuse components of x-ray, neutron or
electron powder diffraction intensity.  For elastic scattering, $Q = 4 \pi
\sin(\theta) / \lambda$, where $\lambda$ is the scattering wavelength and
$2\theta$ is the scattering angle. In practice, values of $\qmin$ and $\qmax$
are determined by the experimental setup and $\qmax$ is often reduced below the
experimental maximum to eliminate noisy data from the PDF since the signal to
noise ratio becomes unfavorable in the high-$Q$ region.

The PDF gives the scaled probability of finding two atoms in a material a
distance $r$ apart and is related to the density of atom pairs in the
material.~\cite{egami;b;utbp03}  For a macroscopic scatterer, $G(r)$ can be
calculated from a known structure model according to
\begin{align}
\label{eq:Grfromrhor}
  G(r) &= 4 \pi r \left[ \rho(r) - \rho_{0} \right], \\
  \rho(r) &= \frac{1}{4 \pi r^{2} N}
            \sum_{i}\sum_{j \neq i}
                \frac{b_{i}b_{j}}{\langle b \rangle ^{2}}
                \delta (r - r_{ij}). \nonumber
\end{align}
Here, $\rho_{0}$ is the atomic number density of the material and $\rho(r)$ is
the atomic pair density, which is the mean weighted density of neighbor atoms at
distance $r$ from an atom at the origin. The sums in $\rho(r)$ run over all
atoms in the sample, $b_{i}$ is the scattering factor of atom $i$, $\langle b
\rangle$ is the average scattering factor and $r_{ij}$ is the distance between
atoms $i$ and $j$.

In practice, we use \eqs{Grfromrhor} to fit the PDF generated from a structure
model to a PDF determined from experiment. For this purpose, the delta
functions in \eqs{Grfromrhor} are Gaussian-broadened and the equation is
modified to account for experimental effects. PDF modeling is performed by
adjusting the parameters of the structure model, such as the lattice constants,
atom positions and anisotropic atomic displacement parameters, to maximize the
agreement between the theoretical and an experimental PDF.  This procedure is
implemented in \PDFgui,~\cite{farro;jpcm07} which is the program used in this
study.  \PDFgui~uses the Levenberg-Marquardt
algorithm~\cite{leven;qam44,marqu;siamjam63} to locally optimize the model
structure.  The algorithm also provides estimates of uncertainties on those
parameters upon convergence, though strictly the estimates are only accurate if
the data are independent and the statistical errors are Gaussian distributed
and properly determined.~\cite{schwa;aca89}

\section{The Nyquist-Shannon sampling theorem}

The Nyquist-Shannon sampling theorem specifies an upper bound on the sampling
interval of a discretized signal in the time domain such that the sample
contains all the available frequency information from the signal.  This upper
bound is $\pi/\Delta\omega$, where $\Delta\omega$ is the angular frequency
bandwidth of the signal.~\cite{shann;pire49} The quantity $\pi/\Delta\omega$ is
commonly referred to as the \emph{Nyquist interval}.  A continuous or discrete
signal sampled on a grid finer than the Nyquist interval can be, in principle,
perfectly reconstructed via interpolation, since the sampling does not
compromise the information content of the signal.

In relation to the PDF, the angular frequency domain is $Q$-space and we are
interested in sampling in $r$-space, the analogue of the time domain. The
frequency information is specified by $F(Q)$ (see \eq{FTofSQtoGr}), which has
bandwidth $\qmax$.%
\footnote{
The sampling theorem as presented in Shannon's paper deals with signals having
positive and negative frequency components. The bandwidth is defined as the
maximum absolute frequency value.  Mathematically, $F(Q)$ is an odd function
(see Eq.~15 in~\cite{farro;aca09}), a fact we use when transforming $F(Q)$ to
$G(r)$ (\eq{FTofSQtoGr}). The ``full'' spectrum of $F(Q)$ that includes the
negative-frequency branch can be calculated from the positive-frequency branch,
and spans the range $[-\qmax, \qmax]$. \qmin does not enter into this since we
enforce $F(Q < \qmin) = 0$ during modeling.~\cite{farro;aca09}
\nba{sjb:make sure this is right.  We do use the negative $Q$ portion!  We
assume that the function is odd and do a sine fourier transform from 0 to
infty, but this is strictly equivalent to an exponential transform from minus
infty to infty } 
\nba{clf:I'm sure of the bandwidth, not the explanation. We don't (can't)
measure data for $Q < 0$, but since $F(Q)$ is odd (in theory) we don't have to.
I've tried to clarify the statement.  }
}
This gives a Nyquist interval of
\begin{equation}
\delrn = \pi/\qmax .
\end{equation}
The sampling theorem states that the PDF can be sampled on any grid with
intervals shorter than this without losing any information from $F(Q)$.


Whittaker~\cite{whitt;prsea15} and Shannon~\cite{shann;pire49} describe an
interpolation formula for reconstructing a signal from samples taken on a grid
with interval, $\delr$, less than the Nyquist interval. In terms of the PDF,
the reconstruction formula is
\begin{equation}
\label{eq:whittpdf}
G'(r) = \sum_n G(n dr) \frac{ \sin(\pi (r/\delr - n) ) }{\pi (r/\delr - n)},
\end{equation}
where $n$ iterates over the points of the sample. Later we will demonstrate the
benefits of modeling the PDF on an optimally sampled grid. This formula allows
us to interpolate a model PDF onto a denser grid, \eg for convenient visual
inspection. In practice, the sampled data must extend beyond the desired range
to avoid reconstruction errors in the high-$r$ region.

\subsection{Aliasing}
\label{sec:aliasing}

Sampling $G(r)$ at or coarser than the Nyquist interval results in
\emph{aliasing}. This term refers to how, in undersampled data, high $Q$
information in $F(Q)$ can masquerade as intensity at lower $Q$.  This is
demonstrated for the PDF by considering its Fourier series over
$-\rmax \leq r \leq \rmax$.
We choose this range because it lets us consider the sine-Fourier series
($G(r)$ is odd) and because the PDF over this range contains the same
information as the PDF over $0 \leq r \leq \rmax$.  Now,
\begin{equation*}
\label{eq:pdfseries}
G(r) = \sum_{m=1}^{\mmax} b_m \sin( Q_m \, r),
\end{equation*}
where $Q_m = m\pi/\rmax$. Since $G(r)$ contains no frequency components greater
than \qmax,
$Q_m \leq \qmax$,
and thus
$\mmax \leq \qmax  \rmax / \pi$.

\begin{figure}
\includegraphics[width=\columnwidth]{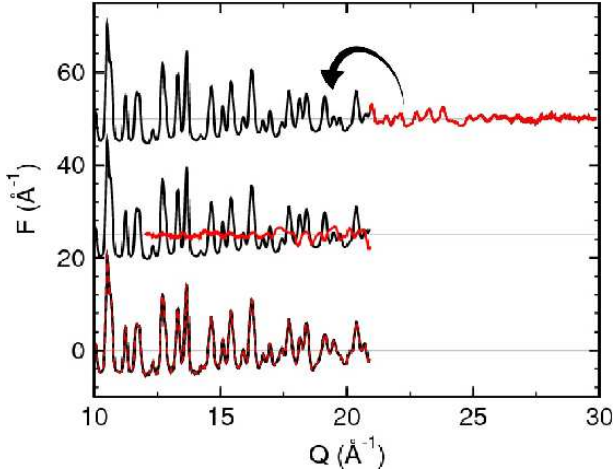}
\caption{
Demonstration of aliasing in $F(Q)$.
(Top) Experimental nickel $F(Q)$ with $\qmax = 29.9~\mAA^{-1}$ featuring
regions above and below $Q' = 20.9~\mAA^{-1}$.
(Center) Experimental nickel $F(Q)$ with the region above $Q'$ ``folded'' over
to lower $Q$.
(Bottom) Aliased $F(Q)$ obtained by sampling the PDF from the experimental
$F(Q)$ on a grid with interval $\delr = 0.15~\mAA$ and Fourier transforming
back to $F(Q)$ (solid line).  This sampling interval is larger than the Nyquist
interval ($\delrn = 0.105~\mAA$) and corresponds to $Q' = \pi / \delr =
20.9~\mAA^{-1}$. Overlaid is the $F(Q)$ obtained by adding the unfolded and
folded segments of the experimental $F(Q)$ (dashed line). Note that the
$Q$-axis starts at $10~\mAA^{-1}$.
}
\label{fig:aliasing}
\end{figure}

Consider the $m^{\mathrm{th}}$ term of the series sampled on the interval
$\delr = \pi / Q'$,
where $Q'$ and $m$ are chosen such that
$Q' \leq Q_m \leq \qmax$.
For the $n^{\mathrm{th}}$ sample, the contribution to the Fourier series is
$ b_m \sin( n \delr \, Q_m)$.
Given the relationship between $Q_m$ and $Q'$,
$n \delr Q_m \geq n (\pi / Q') Q' = n \pi$.
Thus, we can represent the argument as
$(Q_m  - 2Q')\, n \delr + 2n \pi$,
so that the $m^{th}$ frequency component of the sample looks like
$-b_m \sin( (2Q' - Q_m) n \delr)$
for all $n$.  The contribution to $G(r)$ from $F(Q)$ at $Q = Q_m$ therefore
appears in $G(r)$ as if it came from $Q = 2Q' - Q_m$ in $F(Q)$. Consequently,
in $F(Q)$ the signal above $Q'$ gets ``folded'' back to lower $Q$ and overlaps
with the signal in the range
$2Q' - \qmax \leq Q \leq 2Q'$.
This explains how information in $F(Q)$ is progressively lost in $G(r)$ if
$G(r)$ is calculated on grids that are too coarse.  The more undersampled the
data, the greater the $Q$-range that is folded back and the greater the loss of
information in $G(r)$ due to overlapping signals from different $Q$-values.
The effect is illustrated in \fig{aliasing}.
\nba{clf: Would a figure of $G(r)$ and $F(Q)$, both aliased and not help here?}
\nba{sjb: I am still struggling with this description, so yes, a figure might
help.  Maybe I just need to concentrate harder though.  I don't know
intuitively what Q' is for example.}
\nba{clf: $Q'$ is the inverse of the sampling interval. Because of
the discrete nature of the data, and the Fourier transform, we only have so
much information in data. This is measured by \qmax (and the $Q$-spacing). If
we dilute the PDF too much, and transform back to $F(Q)$, then we have
irretrievably lost some of that information. I think the figure helps.}

We note that the case where the data are sampled precisely on a grid with the
Nyquist interval,
$\delr = \delrn$,
then
$Q' = Q_m = \qmax$
and there is no folding. However, there is still loss of information since
$\sin( Q_m n \delr) = 0$,
and so the $m^{th}$ Fourier amplitude, $b_m$, can take on any value.  This is
why a strict inequality between the sampling interval and the Nyquist interval
is required to avoid aliasing:
$\delr < \delrn$.

Aliasing implies that the sampled signal does not uniquely identify its source.
Since some frequency components alias others, the PDF could represent the
aliased $F(Q)$ just as well as the unaliased one. When back-Fourier
transforming a sparsely sampled $G(r)$ into $Q$ space, the aliased $F(Q)$ will
result. The sampling theorem states that aliasing does not occur when sampling
at an interval smaller than the Nyquist interval.

\subsection{Structural Information in the PDF}

\nba{clf:I've rewritten this section to be a bit more cautious, and to reference
a comprehensive Rietveld study.}

The sampling theorem determines the number of data points required to
reconstruct a PDF signal from samples, which is
\begin{equation}
\label{eq:N}
N = \Delta r/\delrn = \frac{\Delta r \qmax}{\pi},
\end{equation}
where $\Delta r$ is the extent of the PDF in $r$-space. What is more relevant
to PDF modeling is the amount of structural information in the PDF. $N$ is an
upper bound on this since we cannot extract more independent observations of
the structure than raw information from the signal. Given perfect data and the
proper model, one can meaningfully extract $N$ structural parameters from a PDF
signal.

Factors such as noise and peak overlap can obscure the structural information
in the PDF and therefore determine whether $N$ is a good estimate of the amount
of structural information in the PDF.  For example, consider a situation where
the PDF contains a single peak, but has a very large \qmax. In this case, a
complete crystal model cannot be obtained from fitting this single peak, no
matter how large $N$ is.  In another extreme case, imagine that the majority of
PDF peaks have a single point or no points due to a small \qmax. In this
situation the anisotropic displacement parameters cannot be determined with
certainty.

In practice, the amount of structural information in the PDF cannot be
precisely known. To perform a reliable refinement, the signal-to-noise ratio
must be favorable,~\cite{egami;b;utbp03} the PDF peaks must be apparent, and
the fit range must be such that the structural features one is seeking to model
are accessible. In addition to this, we recommend using Rietveld refinement
guidelines when refining the PDF, which advise that the ratio of independent
observations to the number of refinement parameters should be around three to
five, preferring the latter.~\cite{mccus;jac99}

\section{Method}
\label{sec:method}

Powder diffraction data were collected from nickel and LaMnO$_3$ (LMO) samples.
The nickel data were collected using the rapid acquisition pair distribution
function (RaPDF) technique~\cite{chupa;jac03} with synchrotron x-rays on
beamline 6-ID-D at the Advanced Photon Source at Argonne National Laboratory.
The sample was purchased from Alfa Aesar. The powdered sample was packed in a
flat plate holder with thickness of 1.0 mm and sealed between Kapton tapes.
Data were collected at room temperature in transmission geometry with an x-ray
energy of 98.001 keV ($\lambda = 0.12651~\mAA$). An image plate camera (Mar345)
with diameter of 345 mm was mounted orthogonally to the beam with a sample to
detector distance of 178.4 mm.

The raw 2D data were reduced to 1D integrated intensity profiles using the
\FittwoD~program.~\cite{hamme;esrf98}
\nba{sjb: what is the monitor?  We don't normally this correction for RAPDF
data, but if we did here we need to describe how the monitor data were
collected.}
\nba{clf: I'll look into it.}
\nba{clf: I assumed this was standard procedure. All I know about the data is
written in
$/u24/masadeh/research/reports/03am_simplElements/smplElmnt.tex$,
which does not mention monitor counts, so I've removed the statement.
}
Corrections for environmental scattering, incoherent and multiple scattering,
polarization and absorption were performed according to the standard
procedures~\cite{egami;b;utbp03} using \PDFgetXtwo~\cite{qiu;jac04i} to obtain
the PDF with $\qmax = 29.9~\mAA^{-1}$.  This corresponds to $\delrn =
0.105~\mAA$.
\nba{sjb: please check the sig figs here, they don't seem to be consistent.
Don't spend too much time on it, but qmax should probably be 30.0 or 30.00
rather than 30, for example, to specify delrn as 0.1047.  Same below.}
\nba{clf:Found chi files, good to 3 sig. figs. nba'd}

\nba{describe how, parameters used etc.}
\nba{clf:more detail?}
\nba{did you use PDFgetN?  are there any other relevant parameters that might
affect the result (I think you have captured the most important ones
below)}
\nba{clf:I don't know who reduced the data. I did a bunch of searching on
slapper for the experiment details, but found little detail about the data
reduction. I've listed the standard stuff. I know even less about the neutron
data, but the reference paper states that PDFgetN was used. The details of the
data reduction are not in the reference.}
\nba{most of the info should be recorded in the header info in the .gr file, if
we have access to that.  What you have is sufficient, though let's add a
reference to PDFgetN.}
\nba{clf:Reference cited.}

The LMO data were collected using time-of-flight neutron diffraction at
the NPDF instrument at the Los Alamos Neutron Scattering Center at Los Alamos
National Laboratory.  The LMO sample preparation and data collection have
been described in detail elsewhere.~\cite{qiu;prl05} The LMO PDFs were
produced with \PDFgetN~\cite{peter;jac00} using $\qmax = 32.0~\mAA^{-1}$. This
corresponds to $\delrn = 0.0982~\mAA$.

\nba{check this is right.}
\nba{clf:This reference seems right given the metadata and the bibs.}
\nba{sjb: OK, I nba'd it.}

In each case, experimental PDFs were generated with $\rmax=20~\mAA$ using
$\delr = 0.01~\mAA$.  PDF data on sparser grids were created by removing points
from this PDF in order to get the desired sampling interval. Pruning the data
in this way is equivalent to recalculating the PDF from $F(Q)$ on the sparser
grid. We produced 31 data-sets with varying \delr against which models were
refined.

We took as a reference data-set the PDF generated on the default grid of
$\delr = 0.01~\mAA$ and structural models were refined to the data.  We then
refined the same models to data-sets on sparser grids.  We define as
$\Delta_p(\delr)$ for a parameter $p$ as the absolute difference between the
value of the parameter $p$ refined for the data-set sampled at interval $\delr$
and that refined for the reference data-set. The accuracy of the refined
parameters becomes unacceptable when $\Delta_p(\delr)$ exceeds the statistical
uncertainty on the difference, $\sigma(\Delta_p(\delr))$.  This is given by
$\sigma(\Delta_p(\delr)) = \sqrt{ \sigma^2(p(\delr)) + \sigma^2(p(0.01))}$,
where $\sigma(p(\delr))$ and $\sigma(p(0.01))$ are the estimated uncertainties
on parameter $p$ taken from the refinement for the data-set sampled at interval
$\delr$ and the reference data-set, respectively.  To determine if a refined
parameter extracted from a sparse data set is accurate, we define a parameter
quality factor, $\qualp(\delr) = \Delta_p(\delr)/\sigma(\Delta_p(\delr))$. If
$\qualp(\delr)$ is less than or equal to one, the parameter value refined from
the data-set sampled at interval $\delr$ is within the expected uncertainty of
the best estimate and is considered accurate. If $\qualp(i)$ is greater than
one, the change in the parameter's value is greater than the expected
uncertainty, and the result is considered unreliable.

The parameter quality measure, $\qualp(i)$, is biased due to a couple of
assumptions. First, by comparing all results with the refinement of the
undiluted data we assume that this refinement gives the best estimate for each
parameter. The validity of this assumption is dependent on the systematic bias
of the refinement results due to the quality of the data and the suitability of
the refinement model. Since this bias is present in the diluted data as well,
its effects should be negligible. Second, we assume that the uncertainty value
derived from the refinement results is accurate. We discuss later that the
uncertainty values derived from refinements of oversampled data-sets are too
small. This inflates the estimated quality factor when the data are
oversampled, but does not invalidate the accompanying results.

The refinements from unaltered and sampled data-sets were performed identically
over a range from $\rmin=0.01~\mAA$ to $\rmax=20.0~\mAA$ using the program
\PDFgui.~\cite{farro;jpcm07}  For the nickel data, the lattice parameter,
isotropic atomic displacement parameter (ADP), dynamic correlation factor,
scale factor and resolution factor were varied in the refinements. In the
LMO fits, three lattice parameters, four isotropic ADPs (one each
for the La, Mn and axial and planar oxygen atoms), and seven fractional
coordinates were varied along with the scale and correlation factors
(see~\cite{bozin;pb06}).  From \eq{N} we get that refinements over this range,
$\Delta r = 19.99~\mAA$, yield $N_{Ni}=191$ and $N_{LMO}=203$.  For the nickel
data set, we have an observation-to-parameter ratio (OPR) greater than 30 and
for LMO the OPR is greater than 10.  The refinements are therefore comfortably
over-constrained and the optimization problem is well conditioned.

\nba{clf: Removed this discussion.}
\nba{sjb: added it back}
\nba{clf: Looks good}

Various refinements were timed to measure the speed-up in the program execution
due to sampling.

\section{Results}
\label{sec:results}

\begin{figure}
\includegraphics[width=0.85\columnwidth]{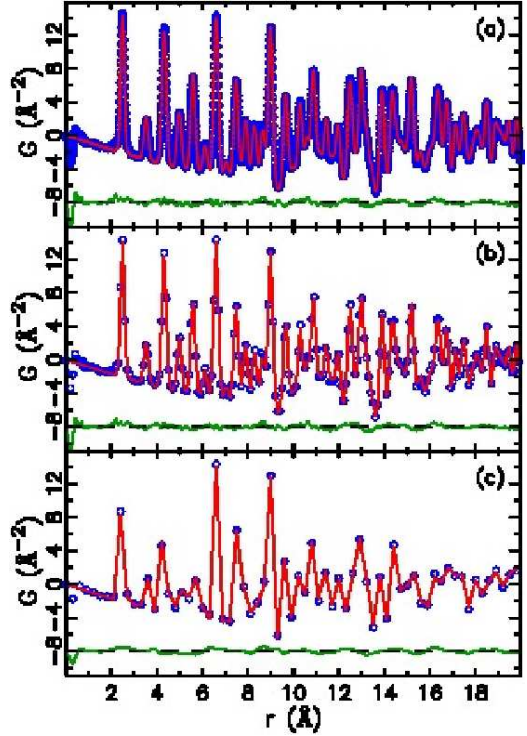}
\caption{
Fits to sampled nickel PDFs.
(a) Unaltered data with $\delr = 0.01~\mAA$.
(b) Sampled data with $\delr = 0.1~\mAA$.
(c) Sampled data with $\delr = 0.3~\mAA$.
The data are shown as circles, the fits are the lines through the data and the
difference is shown offset below.  All fits are of similar quality, despite the
poor visual quality of the data in panels (b) and (c). The data shown in panel
(c) is undersampled and produced unacceptably uncertain results, though this is
not apparent from the difference curve.
}
\label{fig:sparseni}
\end{figure}

\begin{figure}
\includegraphics[width=0.85\columnwidth]{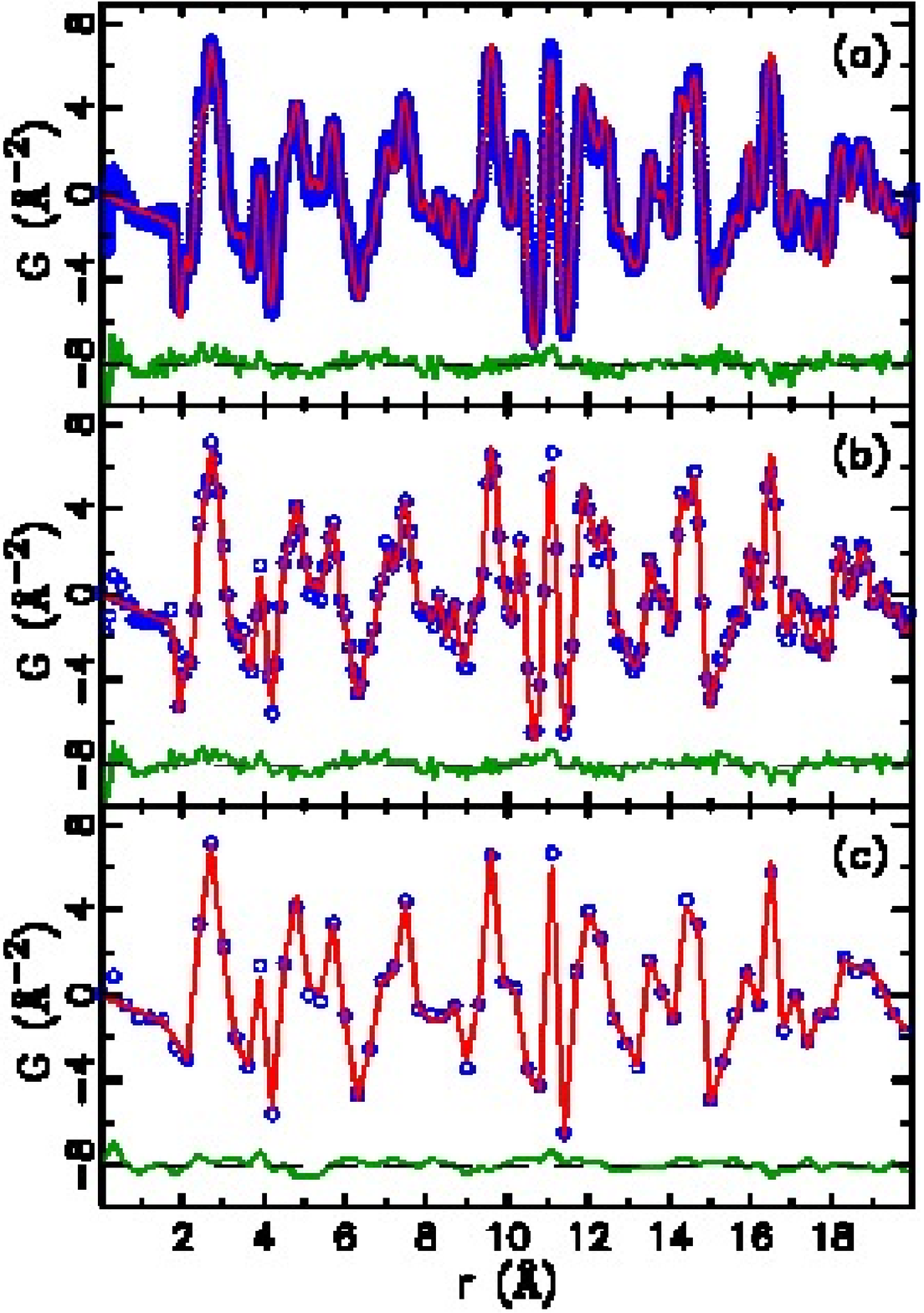}
\caption{
Fits to sampled LaMnO$_3$ PDFs.
(a) Unaltered data with $\delr = 0.01~\mAA$.
(b) Sampled data with $\delr = 0.1~\mAA$.
(c) Sampled data with $\delr = 0.3~\mAA$.
The data are shown as circles, the fits are the lines through the data and the
difference is shown offset below.  All fits are of similar quality, despite the
poor visual quality of the data in panels (b) and (c). The data shown in panel
(b) and (c) are undersampled, and the data in panel (c) produced unacceptably
uncertain results. Note that in panel (c) several peaks are not resolved.
}
\label{fig:sparselmo}
\end{figure}

When the nickel and LMO data are made sparser, the PDF profiles appear less
smooth and the detailed shape of the peak profiles becomes less apparent.  This
is shown in \figs{sparseni} and \ref{fig:sparselmo}. The data in panel (a) in
both figures are on the reference grid ($\delr = 0.01~\mAA$) and are both
smooth and have well-defined Gaussian-like peaks.~\cite{egami;b;utbp03} The
data in panel (b) are sampled with $\delr = 0.1~\mAA$, close to the Nyquist
interval, and are not nearly as smooth, though the peaks are still well
defined.  Lastly, the data in panel (c) are sampled with $\delr = 0.3~\mAA$,
where there is apparent loss of information.  The refined parameters from these
fits are given in \tabl{niresults} and \tabl{lmoresults}.  Note that the
uncertainty in the refined parameters increases from $\delr = 0.01~\mAA$ to
$\delr = 0.1~\mAA$, although each of these data-sets produce acceptable
results.
\begin{table}
\caption{
Parameters from Ni refinements using data with various \delr. The Nyquist
interval, \delrn, is $0.105~\mAA$. Here, $a$ denotes the lattice parameter,
$U_{iso}$ the isotropic ADP, $\delta_2$ the vibrational correlation parameter,
$scale$ the data scale and $Q_{damp}$ the experimental resolution factor.
}
\label{table:niresults}
\begin{tabular}{p{0.19\columnwidth}p{0.19\columnwidth}
p{0.19\columnwidth}p{0.19\columnwidth}p{0.19\columnwidth}}
    \hline
               $\delr (\mAA)$
            &  $0.01$
            &  $0.10$
            &  $0.12$
            &  $0.30$  \\
    \hline
    $a (\mAA)$
        & 3.53159(2)    & 3.53158(6)    & 3.53158(6)    & 3.53186(10)   \\
    $U_{iso} (\mAA^2)$
        & 0.005446(7)   & 0.00545(2)    & 0.00543(2)    & 0.00570(4)    \\
    $\delta_2 (\mAA^2)$
        & 2.25(2)       & 2.20(5)       & 2.15(5)       & 2.2(2)        \\
    $scale$
        & 0.7324(7)     & 0.733(2)      & 0.734(3)      & 0.761(4)      \\
    $Q_{damp} (\mAA^{-1})$
        & 0.06307(11)   & 0.0632(4)     & 0.0634(4)     & 0.0653(7)     \\
    \hline
\end{tabular}
\end{table}
\begin{table}
\caption{
Parameters from LaMnO$_3$ refinements using data with various \delr.  The
Nyquist interval, \delrn, is $0.0982~\mAA$.  Here, $a$, $b$ and $c$ denote the
lattice parameters, $U_{iso}$ the isotropic ADP (one for each primitive atom),
$x$, $y$ and $z$ the fractional atomic coordinates, $\delta_2$ the vibrational
correlation parameter and $scale$ the data scale.
}
\label{table:lmoresults}
\begin{tabular}{p{0.19\columnwidth}p{0.19\columnwidth}
p{0.19\columnwidth}p{0.19\columnwidth}p{0.19\columnwidth}}
    \hline
               $\delr (\mAA)$
            &  $0.01$
            &  $0.10$
            &  $0.12$
            &  $0.30$  \\
    \hline
    $a (\mAA)$
        & 5.5394(2)     & 5.5394(6)     & 5.5393(7)     & 5.5362(14)    \\
    $b (\mAA)$
        & 5.7441(2)     & 5.7443(7)     & 5.7442(8)     & 5.7536(13)    \\
    $c (\mAA)$
        & 7.7059(2)     & 7.7059(9)     & 7.7054(10)    & 7.697(2)      \\
    $\delta_2 (\mAA^2)$
        & 2.44(3)       & 2.38(9)       & 2.35(9)       & 2.49(14)      \\
    $scale$
        & 0.7941(11)    & 0.794(3)      & 0.795(4)      & 0.803(6)      \\
    \multicolumn{5}{l}{\textbf{La}} \\
    $x$
        & 0.99234(10)   & 0.9923(3)     & 0.9926(4)     & 0.9917(6)     \\
    $y$
        & 0.04828(8)    & 0.0482(2)     & 0.0481(3)     & 0.0469(5)     \\
    $U_{iso} (\mAA^2)$
        & 0.00508(4)    & 0.00506(13)   & 0.0052(2)     & 0.0055(2)     \\
    \multicolumn{5}{l}{\textbf{Mn}} \\
    $U_{iso} (\mAA^2)$
        & 0.00376(7)    & 0.0038(2)     & 0.0038(2)     & 0.0024(3)     \\
    \multicolumn{5}{l}{\textbf{O$_1$}} \\
    $x$
        & 0.07300(11)   & 0.0730(4)     & 0.0730(4)     & 0.0739(7)     \\
    $y$
        & 0.48625(10)   & 0.4862(3)     & 0.4864(4)     & 0.4874(7)     \\
    $U_{iso} (\mAA^2)$
        & 0.00682(8)    & 0.0067(3)     & 0.0068(3)     & 0.0075(3)     \\
    \multicolumn{5}{l}{\textbf{O$_2$}} \\
    $x$
        & 0.72515(8)    & 0.7251(2)     & 0.7252(3)    & 0.7247(5)      \\
    $y$
        & 0.30682(8)    & 0.3068(3)     & 0.3069(3)    & 0.3072(5)      \\
    $z$
        & 0.03876(6)    & 0.0388(2)     & 0.0389(2)    & 0.0399(3)      \\
    $U_{iso} (\mAA^2)$
        & 0.00689(4)    & 0.0069(2)     & 0.0068(2)    & 0.0062(2)      \\
    \hline
\end{tabular}
\end{table}

\begin{figure}
\includegraphics[width=0.9\columnwidth]{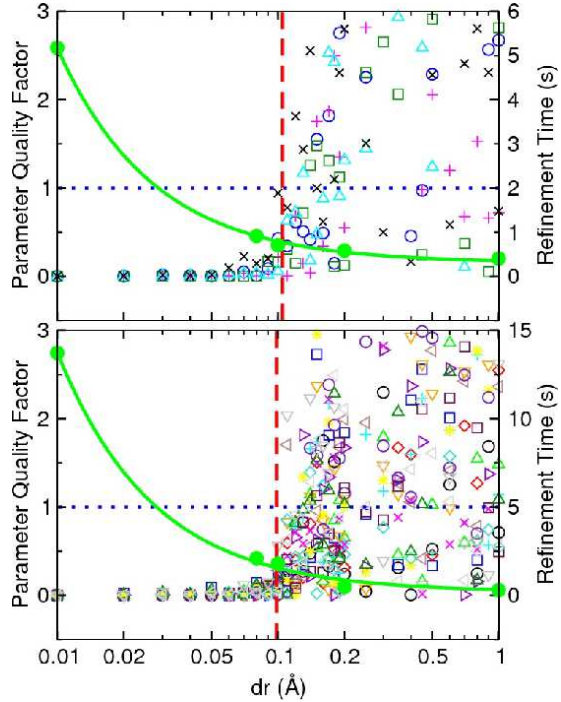}
\caption{
Refined parameter quality (open symbols) and refinement times (solid circles)
measured using sampled Ni (top) and LaMnO$_3$ (bottom) data. The dotted
horizontal line shows the cutoff between acceptable and unacceptable parameter
quality. The dashed vertical line shows the value of \delrn predicted by the
sampling theorem.  For \delr values larger than this the quality of some
parameters transition into the unacceptable region. The time values demonstrate
the decrease in refinement time with increasing \delr, with more than a
seven-fold speed up near \delrn. The solid curve through the time values is fit
to the form $a + b/\delr$.
}
\label{fig:paramresults}
\end{figure}

In \fig{paramresults} we show the parameter quality values, $\qualp(i)$,
plotted against the sampling interval.  The quality factor is satisfactory for
data-sets that are sampled with grids close to the reference data-set.  This
indicates that these refinements are producing the same parameter values.  Near
the Nyquist interval (indicated by the vertical dashed line), various quality
factors become unacceptable.

\section{Discussion}

Figure \ref{fig:paramresults} indicates that the onset of unreliable
refinements coincides with the Nyquist interval. The refined parameter values
are all acceptable, and largely independent of the sampling interval in the
oversampling region ($\delr < \delrn$).  Figures~\ref{fig:sparseni},
\ref{fig:sparselmo} and~\ref{fig:paramresults} indicate that visual appearance
is not a good indicator of data quality.

The sampling theorem tells us that the information content in the data does not
change as long as we sample on a grid finer than the Nyquist interval.  We
expect to and do refine the same parameters from such samples.  As the data are
sampled onto grids coarser than the Nyquist interval, we expect to lose
structural information gradually.  In contrast, refined values of the
parameters become unreliable quickly as the Nyquist interval is exceeded.  This
is somewhat surprising since the refinements are highly overconstrained and
have an estimated OPR greater than 5 even when sampled at twice the Nyquist
interval. In \fig{paramresults} we see the quality of the refined parameters
diverge well before this point.  Intuition would tell us that it is possible to
lose a considerable quantity of information by sampling before refinements
become unstable.  This is not observed.  The instability is not caused solely
by information loss, but by information corruption due to aliasing.

Aliasing has two effects on a PDF signal, as described in \sect{aliasing}.
Foremost, aliasing lowers the effective maximum $Q$-value in $F(Q)$ from \qmax
to $Q' = \pi / \delr$. This creates the obvious effect of lower resolution in
the PDF, as seen in \figs{sparseni} and \ref{fig:sparselmo}. In extreme cases,
this will lead to poorly defined peaks in the PDF.  Less obviously, sampling on
a grid coarser than the Nyquist interval allows for the possibility that the
PDF has originated from a different, aliased, $F(Q)$ as shown in
\fig{aliasing}. When calculating the model PDF, we enforce $F(Q > \qmax) = 0$.
When there is aliasing the structure function resulting from $G(r)$ has $F(Q >
\pi / \delr = 0)$, and extra intensity below $\pi / \delr$.  Thus, aliasing
makes it possible to find a different set of refinement parameters that
describes the sampled PDF.  This is true regardless of the optimization
algorithm.

The estimated uncertainties on the fitting parameters for \delr in the region
of stable refinements are dependent on the sampling interval.  We see from
Tables \ref{table:niresults} and \ref{table:lmoresults} that the uncertainties
on the parameters increase when estimated from the data sampled near the
Nyquist interval compared to the reference data.  The sampling theorem gives
the number of data points necessary to fully represent the PDF.  Any data
sampled on a grid finer than the Nyquist interval are necessarily redundant.
If a set of fitting parameters reproduces a particular set of points well on a
optimal grid, those parameters will also reproduce the associated redundant
points well.  By not taking into account the correlations between data
points,~\cite{toby;aca04} as in this study, this results in under-estimated
uncertainty values on parameters.  Refining optimally sampled data reduces
these correlations while retaining all the structural information available in
the data and gives a more reliable estimate of uncertainties.

A fortunate side-effect of refining optimally sampled data is a decreased
refinement time.  Shown in \fig{paramresults} is a plot of refinement times for
some chosen sampling intervals.  The trend in the plot shows that refinement
time is proportional to the inverse of \delr (shown as the broad solid line),
or directly proportional to the number of data points, with a constant offset.
This trend reflects the fact that the calculation of the PDF grows linearly
with the number of sample points.  Carrying out refinements on optimally
sampled data gives a significant speed increase compared to the reference data;
in this case the speed increases by more than a factor of seven.

These observations indicate that PDF refinements should be performed on the
sparsest grid possible with sampling interval less than the Nyquist interval.
To produce an esthetically pleasing presentation of the PDF, one can
interpolate onto a finer grid using the Whittaker-Shannon interpolation formula
(\eq{whittpdf}).

\nba{clf: I've removed the discussion of fitting data over limited ranges
because I'm not sure how well it fits now. This is covered somewhat in new
``amount of information'' section. I have no objections to putting it in the
manuscript, but I wasn't sure how to do that given the new information.
}

\nba{clf: I've removed all mention of the degrees of freedom. I'm not exactly
sure where it fits in. We can get unreliable fitting parameters by
over-parameterizing a refinement. I don't think this is orthogonal to aliasing,
but I don't know how they are related.

We expect refinements to become unstable when the number of refined independent
parameters exceeds $N$.  In this case, $N$ is given by Eq.~\ref{eq:N} for
data-grids smaller $\delrn$, and is the number of data-points for grids that
are coarser than that
\nba{finish discussion....}
\nba{discussion about how estimated precision on refinements should go up for
fits on grids below the SN frequency, but the refinements shouldn't become
unstable until we exceed the dof's.}
}
\nba{sjb: I don't know how to handle this, except to say I don't really
understand why the refinements get unstable so quickly.  I think the DOF
arguments are still valid (and the "aliasing" arguments are not), but the
system is totally not behaving the way I would expect.  There must be something
more to the aliasing errors that drive the fits unstable, but what is it?  Or
is it that we are not estimating the uncertainties correctly?
}

\section{Conclusions}

The purpose of this research was to demonstrate the consequences of the
Nyquist-Shannon sampling theorem as it applies to the PDF.  We show that the
quality of refined parameters diverges when sampling the PDF at intervals
larger than the Nyquist interval, which is the result of aliasing. Furthermore,
we show that the estimated uncertainties of refined parameters are more
reliable when the PDF is optimally sampled.  Statistically reliable
uncertainties on refined parameters can be obtained by taking into account the
correlations between all the points in $G(r)$,~\cite{toby;aca04} but this comes
at the computational expense of inverting a large error matrix. By optimally
sampling the PDF, the correlations among points in the PDF are minimized while
preserving all the available structural information. This gives improved
uncertainty estimates without costly computation, and may expedite refinements
when the PDF can be computed over fewer points.

The Nyquist-Shannon sampling theorem gives an upper bound on the amount of
structural information contained in an experimental PDF. This determines the
$Q$- and $r$-extent that are required for a model refinement to be
overconstrained.  Oversampling the PDF does not add more information to a
refinement, and therefore provides no benefit other than an esthetically
pleasing visualization.  This result emphasizes the importance of collecting
diffraction data to high $Q$ when it is to be used for PDF modeling, since a
larger $\qmax$ decreases the Nyquist interval, and makes accessible smaller
structural details.

\section{Acknowledgments}

\nba{clf:A reviewer suggested we remove Margaret's acknowledgments, I'll leave
that up to you.}

M.~S.~would like to thank the entire Billinge-group lab for helping and
supporting her throughout this summer project - Jiwu, Pavol, Ahmad, and
especially Hyun-Jeong.
Lastly, she thanks Dr. Richmond at Michigan State University for putting
together an amazing summer high-school research experience program.  Research
in the Billinge group was supported by the US National Science foundation
through Grant DMR-0703940. Use of the APS is supported by the U.S. DOE, Office
of Science, Office of Basic Energy Sciences, under Contract No.
W-31-109-Eng-38. The 6ID-D beamline in the MUCAT sector at the APS is supported
by the U.S. DOE, Office of Science, Office of Basic Energy Sciences, through
the Ames Laboratory under Contract No.  W-7405-Eng-82. Beamtime on NPDF at
Lujan Center at Los Alamos National Laboratory was funded under DOE Contract
No. DEAC52-06NA25396.


\end{document}